# Quantifying the Dzyaloshinskii-Moriya Interaction Induced by the Bulk Magnetic Asymmetry

Qihan Zhang[1,2,*], Jinghua Liang[3,*], Kaiqi Bi[4,6,*], Le Zhao[1,2], He Bai[5], Qirui Cui[3],

Heng-An Zhou[1,2], Hao Bai[1,2], Hongmei Feng[1,2], Wenjie Song[6], Guozhi Chai[6],

O. Gladii[7], H. Schultheiss[7], Tao Zhu[5,8], Junwei Zhang[4,6],

Yong Peng[4,6,†], Hongxin Yang[3,9,†], Wanjun Jiang[1,2,†]

[1] *State Key Laboratory of Low-Dimensional Quantum Physics and Department of Physics, Tsinghua University, Beijing 100084, China*

[2] *Frontier Science Center for Quantum Information, Tsinghua University, Beijing 100084, China*

[3] *Ningbo Institute of Materials Technology and Engineering, Chinese Academy of Sciences, Ningbo 315201, China*

[4] *School of Materials and Energy, Electron Microscopy Centre of Lanzhou University, Lanzhou University, Lanzhou 730000, China*

[5] *Spallation Neutron Source Science Center, Dongguan, 523803, China*

[6] *Key Laboratory of Magnetism and Magnetic Materials of Ministry of Education, Lanzhou University, Lanzhou 730000, China*

[7] *Institut für Ionenstrahlphysik und Materialforschung, Helmholtz-Zentrum Dresden-Rossendorf, D-01328 Dresden, Germany*

[8] *Beijing National Laboratory for Condensed Matter Physics and Institute of Physics, Chinese Academy of Sciences, Beijing 100190, China*

[9] *Center of Materials Science and Optoelectronics Engineering, University of Chinese Academy of Sciences, Beijing 100049, China*

[*] These authors contributed equally to this work.

[†] To whom correspondence should be addressed: pengy@lzu.edu.cn

hongxin.yang@nimte.ac.cn

jiang_lab@tsinghua.edu.cn

**Abstract**

A broken interfacial inversion symmetry in ultrathin ferromagnet/heavy metal (FM/HM) bilayers is generally believed to be a prerequisite for accommodating the Dzyaloshinskii-Moriya interaction (DMI) and for stabilizing chiral spin textures. In these bilayers, the strength of the DMI decays as the thickness of the FM layer increases and vanishes around a few nanometers. In the present study, through synthesizing relatively thick films of compositions CoPt or FePt, CoCu or FeCu, FeGd and FeNi, contributions to DMI from the composition gradient induced bulk magnetic asymmetry (BMA) and spin-orbit coupling (SOC) are systematically examined. Using Brillouin light scattering spectroscopy, both the sign and amplitude of DMI in films with controllable direction and strength of BMA, in the presence and absence of SOC are experimentally studied. In particular, we show that a sizable amplitude of DMI ($\pm 0.15$ mJ/m$^2$) can be realized in CoPt or FePt films with BMA and strong SOC, whereas negligible DMI strengths are observed in other thick films with BMA but without significant SOC. The pivotal roles of BMA and SOC are further examined based on the three-site Fert-Lévy model and first-principles calculations. It is expected that our findings may help to further understand the origin of chiral magnetism and to design novel non-collinear spin textures.

**Introduction**

In ultrathin ferromagnet/heavy metal (FM/HM) bilayers, an interfacial asymmetry combined with a strong spin-orbit coupling (SOC) establishes an interfacial Dzyaloshinskii-Moriya interaction (DMI) that gives rise to exciting spin-orbitronic physics[1-6]. Some examples are the observation of chiral domain walls[7-11], Néel skyrmions[12-16], and the driven motion of magnetic domains/skyrmions by spin torques[7, 9, 13, 15, 17, 18]. The SOC strength is not a readily adjustable material parameter, but symmetry breaking can be introduced by different means[19-24]. For example, early work in GdFeCo amorphous films revealed how a naturally existing composition gradient along the thickness direction introduces a bulk DMI[20]. However, a clear mechanistic explanation and more importantly, successful control of both the sign and amplitude of DMI induced by the composition gradient-induced bulk magnetic asymmetry (BMA) are not established. Note that BMA in $Fe_{1-x}Pt_x$ films could lead to bulk spin torques, which result in a self-switching[23, 25-27].

Based on the three-site Fert-Lévy model[28-30], the occurrence of DMI between neighboring magnetic atoms (sites $i$ and $j$) is enabled by the SOC of the HM atom (site $l$) via conduction electrons. The resultant DM vector $D_{ijl}$ reads as[28]:

$$\vec{D}_{ijl}(\vec{R}_{li}, \vec{R}_{lj}, \vec{R}_{ij}) = -V_1 \frac{(\vec{R}_{li} \cdot \vec{R}_{lj})(\vec{R}_{li} \times \vec{R}_{lj})}{|\vec{R}_{li}|^3 |\vec{R}_{lj}|^3 |\vec{R}_{ij}|}, \quad (1)$$

where $\vec{R}_{li}$, $\vec{R}_{lj}$, and $\vec{R}_{ij}$ are the corresponding distance vectors, $V_1$ is a SOC-governed material parameter. For thick films without BMA, DM vectors ($\vec{D}_{ijl}$ and $\vec{D}_{ijl'}$) on the opposite sides of the FM atoms are compensated, as shown in Fig. 1(a). In the presence of BMA, finite, uncompensated DM vectors can be found, as shown in Figs. 1(b) and 1(c), respectively. In particular, the sign of the uncompensated DM vector is determined by the direction of BMA. In the present work, both the role of BMA and SOC for producing DMI will be experimentally examined.

**Synthesis of thick films with bulk magnetic asymmetry**

The presence of BMA is introduced by varying the relative composition of binary alloys ($FM_xNM_{1-x}$) along with the thickness ($t$) direction, where FM denotes the magnetic element (Co, and Fe) and NM is the non-magnetic metal with (or without) SOC. We define the starting ($x_i$)/ending ($x_t$) compositions as $FM_{x_{i,t}}NM_{1-x_{i,t}}$. The composition gradient reads as $\Delta x/t = (x_t - x_i)/t$, where $\Delta x = x_t - x_i$ is the composition difference between the ending and starting layers. The corresponding magnetization gradient reads as $\Delta M_s/t = (M_s^t - M_s^i)/t$, where $M_s^{i,t}$ are the saturation magnetization of the starting and ending layers, respectively.

Binary magnetic films with in-plane magnetic anisotropy are grown using the co-sputtering technique. During the growth, the relative deposition rates of the FM and NM elements are linearly changed, which create a linear composition/magnetization difference along the growth direction. Multilayers of stacking order Ta(3 nm)/$FM_xNM_{1-x}$($t$ nm)/Ta(3 nm) are synthesized, in which both the choice of FM/NM elements, the direction of composition difference ($\pm\Delta x$), the effect of magnetization differences ($\Delta M_s/\Delta x$) and the thickness ($t$) will be studied. A full list of samples, together with the key parameters are listed in Table S1 in part S2 of the Supplemental Materials[31].

The presence of composition differences is examined by using a high-angle annular dark-field scanning transmission electron microscopy (HAADF-STEM). HAADF-STEM images obtained from two representative samples, CoPt ($\Delta x = \pm 52\%$), are shown in Figs. 2(a) and 2(e). The opposite contrast changes within the images are associated with the opposite composition differences, which are further verified using energy-dispersive X-ray spectroscopy (EDS), as shown in Figs. 2(b) and 2(f), respectively. The composition gradients are also studied in other films (see part S5 of Supplemental Materials[31]).

To correlate the composition and magnetization gradients, we conducted polarized neutron reflectometry (PNR) measurement (see the Supplemental Materials[31]). From the nuclear scattering length density, opposite trends of composition from $Co_{0.25}Pt_{0.75}$ to $Co_{0.63}Pt_{0.37}$ (for $\Delta x = +52\%$) and from $Co_{0.63}Pt_{0.37}$ to $Co_{0.25}Pt_{0.75}$ (for $\Delta x = -52\%$) are revealed. Based on the profile of magnetic scattering length density, the corresponding magnetization gradients are further estimated, as shown in Figs. 2(c) and 2(g), respectively. From the PNR measurement, values of $M_s$ for $Co_{0.25}Pt_{0.75}$ and $Co_{0.63}Pt_{0.37}$ are determined to be $0.453 \times 10^6$ and $1.082 \times 10^6$ A/m. By using a quasilinear approximation, the ratio of $\Delta M_s/\Delta x = 1.655 \times 10^6$ A/m is determined.

A monotonic decrease of $M_s$, following the increase of $\Delta x$, is measured by vibrating sample magnetometry (VSM), as shown in Fig. 2(d). For samples with opposite composition gradients ($\pm\Delta x$), values of $M_s$ remain approximately the same. For films with homogenous compositions ($\Delta x = 0$), a linear dependence of $M_s$ as a function of homogenous composition ($x$) is obtained for $Co_xPt_{1-x}$ and $Fe_xPt_{1-x}$ films, as shown in Fig. 2(h). The linear slope of $M_s(x)$ is calculated, which enables the ratio of $\Delta M_s/\Delta x$ (in the unit of $10^6$ A/m) to be obtained: 1.67 for CoPt ($\Delta x$), 2.09 for FePt ($\Delta x$), 1.31 for CoCu ($\Delta x$) and 1.78 for FeCu ($\Delta x$), respectively. These values are in agreement with the PNR results. Upon establishing a quasilinear relation of $\Delta M_s/\Delta x$, we use $\Delta x$ to represent the magnetization gradient ($\Delta M_s/t$, with $t = 6$ nm).

**Measurement of the strength of DMI induced by BMA**

By using Brillouin light scattering (BLS) spectroscopy, both the amplitude and sign of DMI can be determined through measuring the non-reciprocal frequency shift of Damon–Eshbach (DE) spin waves[44-48], as schematically illustrated in Fig. 3(a). The dispersion of the DE spin waves reads as[48, 49],

$$f(\mathbf{k}) = \frac{\gamma}{2\pi}\sqrt{(H + J\mathbf{k}^2)(H + J\mathbf{k}^2 + \mu_0 M_{\text{eff}})} - \text{sgn}(M_z)\frac{\gamma}{\pi M_s}Dk_x \quad (2)$$

where $J = 2A/M_s$, with $A$ being the exchange constant, $D$ the volume-averaged DMI constant, $\gamma$ the gyromagnetic ratio, and $\mu_0 M_{\text{eff}}$ the effective demagnetization field. The projection of the spin-wave vector ($\mathbf{k}$) in the x-direction $k_x = \frac{4\pi}{\lambda}\sin\theta$ is determined by the incident angle ($\theta$) and the wavelength of the laser ($\lambda = 532$ nm). The frequency shift ($\Delta f$) between the counterpropagating spin waves ($\pm k_x$) induced by DMI is given by:

$$\Delta f = f_{DM}(-k_x, M_z) - f_{DM}(k_x, M_z) = \frac{2\gamma}{\pi M_s}Dk_x \quad (3)$$

Thus, $\Delta f$ between the Stokes/anti-Stokes peaks measured at negative and positive fields enables $D$ to be quantified, through a linear fitting of $\Delta f$ vs. $k_x$.

Typical BLS spectra for CoPt ($\Delta x = \pm 52\%$), with $k_x = 11.8$ rad/μm, are shown in Figs. 3(b) and 3(c). In CoPt ($\Delta x = +52\%$) film, one observes a positive frequency shift ($+\Delta f$) under opposite magnetic fields ($\mu_0 H = \pm 200$ mT), as shown in Fig. 3(b). In CoPt ($\Delta x = -52\%$) film, there, however, exhibits a negative frequency shift ($-\Delta f$), as shown in Fig. 3(c). These opposite frequency shifts ($\pm\Delta f$) imply an opposite sign of DMI in samples with opposite magnetization gradients ($\pm\Delta x$), as suggested by Eq. (3). Figs. 3(d) and 3(e) summarize the evolution of $\Delta f\ vs.\ k_x$ for CoPt and FePt films, respectively. Following the increase of $\Delta x$, the slope of $\Delta f\ vs.\ k_x$ increases monotonically. Upon inverting the sign of $\pm\Delta x$, the reversed signs of the slopes suggest that the sign of DMI is related to the direction of BMA. Note that a precise determination of the volume-averaged DMI constant $D$ will be discussed later. A small offset of $\Delta f$ at $k_x = 0$ from the origin is observed, which may arise from the different magnetic properties of the top/bottom surfaces that naturally lead to an additional asymmetry of the counter-propagating (dipolar) spin waves[50].

To elucidate the necessity of SOC and BMA in producing DMI, three types of reference samples are examined. The first set of samples is $Fe_{1-x}Pt_x$ and $Co_{1-x}Pt_x$ films with homogeneous compositions ($x$). These samples contain HMs (and have significant SOC) but have no magnetization gradients and hence no BMA ($\Delta x = 0$). The second set of samples is CoCu ($\Delta x$), FeCu ($\Delta x$) and FeNi ($\Delta x$) films. These samples have magnetization gradients ($\Delta x \neq 0$) but no HMs (and have insignificant SOC). The third set of samples is FeGd ($\Delta x$) films to clarify whether the SOC from the 4f rare-earth metal can also establish a bulk DMI, in the presence of BMA.

In the first type of films ($\Delta x = 0$, $t = 6$ nm), we observe a nearly zero slope from $\Delta f\ vs.\ k_x$, indicating a negligible DMI (see part S6 of Supplemental Material[31]). This observation can be explained by the compensated DMI vectors from the equivalent probability of Pt atoms that appear at the nearest Co/Fe sites, as implied by Fig. 1(a). In the other two sets of samples, the role of SOC is examined via replacing Pt with Cu which exhibits a negligible SOC, or with 4f rare earth metal Gd. The BLS spectra of CoCu ($\Delta x = +51\%$) film at $k_x = 11.8$ rad/μm and under $\mu_0 H = \pm 200$ mT are shown in Fig. 3(f), in which the absence of frequency shift is observed. A negligible frequency shift ($\Delta f$) is also observed in FeGd ($\Delta x = +37\%$) films at $k_x = 16.7$ rad/μm and under $\mu_0 H = \pm 1000$ mT, as shown in Fig. 3(g). This result indicates the SOC of the Gd cannot establish a bulk DMI in the FeGd binary films with in-plane anisotropy. The intriguing origin of the bulk DMI observed in GdFeCo and similar compounds[20] may require a revisit. Through identifying a nearly zero slope of $\Delta f$ vs. $k_x$, vanishingly small DMI values are summarized in Fig. 3(h). Note that a vanishingly small DMI in FeNi ($\Delta x = \pm 48\%$) films are also obtained (see part S6 of Supplemental Material[31]). These experiments suggest that both BMA and SOC from 5d HM are co-requisites for producing sizable DMI in thick films, as suggested by the Fert-Lévy model.

Summarized in Fig. 4 (a) is the volume-averaged DMI constants $D$. For CoPt ($\Delta x$) and FePt ($\Delta x$) films, the strength of DMI increases with the increase of the magnetization gradient ($\Delta x$), and the polarity of DMI flips upon inverting $\pm \Delta x$. For CoCu ($\Delta x$), FeCu ($\Delta x$), FeGd ($\Delta x$) and FeNi ($\Delta x$) films, negligible DMI constants are consistent with the absence of SOC from 5d HM. These results further confirm that the non-reciprocity of spin waves originating from dipolar fields[51], interface anisotropies[52] and different values of $M_S$ between the top and bottom magnetic layers are not contributing factors. Note that several SiN(5 nm)/FM$_x$NM$_{1-x}$($t$ nm)/SiN(5 nm) films are also made, in which the same sign and approximately same amplitude of DMI are observed, as shown in Fig.13 of the Supplemental Materials[31]. This result is expected from the negligible (compensated) DMI from the dual interfaces in Ta/FM$_x$NM$_{1-x}$/Ta trilayer.

By fixing the composition of the starting/ending layers [CoPt ($\Delta x = +28\%$) and FePt ($\Delta x = +19\%$)], the BMA induced DMI is further examined through varying the thickness from 3 nm to 9 nm. In this case, the values of $M_S^{i,t}$ of the starting/ending layers are fixed. In particular, the linear evolution of $D$ vs. $\Delta x/t$ can be found in Fig. 4(b), confirming the key role of BMA in determining the strength of DMI.

**The theoretical formalism of the BMA-induced DMI**

Based on the three-site Fert-Lévy model and the first-principles calculations, the BMA-

induced DMI is theoretically examined. Specifically, the total DMI energy ($E_{DM}$) can be written as[40]:

$$E_{DM} = \sum_{i,j,l} \vec{D}_{ijl} \cdot (\vec{S}_i \times \vec{S}_j). \tag{4}$$

For different starting compositions $x_i$, with the number of layers $L = 100$, the calculated $E_{DM}$ $vs.\Delta x$ in the framework of the three-site Fert-Lévy model are shown in Fig. 5(a). A monotonical increase of $E_{DM}$ $vs.\Delta x$, together with a reversed sign of $E_{DM}$ by flipping $\pm\Delta x$ can be observed. In particular, $E_{DM}$ vanishes at $\Delta x = 0$ (without a composition gradient). Upon fixing the composition of the starting/ending layers, the evolution of DMI strength as a function of thickness is also calculated and shown in Fig. 5(b). There, one can find that the calculated $E_{DM}$ monotonically increases with the increased gradient ($\Delta x/L$) under fixed composition differences ($\Delta x$ =+10%, +20%, +30%, +40% and +50%). This is consistent with our experimental observations shown in Fig. 4 (b).

To quantify the strength of BMA-induced DMI, first-principles calculations are also performed (see Supplemental Material[31]). We consider an hcp stacking with lattice parameters being $a$ and $c = 1.6a$, which consists of $100 \times 100 \times 0.5L$ hexagonal unit cells and calculate the $E_{DM}$ of a cycloidal spin spiral with a small spiral vector $q = (2\pi/100a, 0, 0)$. For example, the structural configurations with $\Delta x = \pm 42.85\%$ are shown in the insets of Fig. 5(c). The calculated DMI constants ($d_{DM}$) of the 4-layers CoPt thin films with different $\Delta x$ are shown in Fig. 5(c) in which a non-vanishing DMI for all $\Delta x \neq 0$ films can be identified. Both the sign and magnitude of DMI match with the experimental results (as shown in Fig. 4(a)). These observations demonstrate that the strength of DMI is directly determined by the amplitude of magnetization gradient ($\Delta x$), and the polarity of which is determined by the sign of $\pm\Delta x$.

**Conclusions**

The volume-averaged DMI is comprehensively investigated through BLS spectroscopy in CoPt or FePt, CoCu or FeCu, NiFe and FeGd magnetic films with magnetization/composition gradients ($\Delta x$) along the growth direction. Through measuring the amplitude and sign of DMI, a connection between the sign/amplitude of magnetization/composition gradients ($\pm\Delta x$), and the strong SOC is obtained. In particular, a relatively large DMI value, $\pm 0.15$ mJ/m$^2$, is obtained in 6 nm thick CoPt films with $\Delta x = \pm 52\%$, whereas the absence of DMI in other films without involving BMA and SOC is revealed. The evolution of the BMA-induced DMI is theoretically examined through performing atomic calculations, which agree with the experimental observations. The identification of the bulk DMI that is jointly produced by BMA and SOC of the 5d HM, could serve as an important ingredient for the current understandings of chiral magnetism, and could help to design novel chiral non-collinear spin textures[53].

**Acknowledgments**. Work carried out at Tsinghua was supported by the Basic Science Center Project of NSFC (Grant No. 51788104), National Key R&D Program of China (Grant Nos. 2017YFA0206200), the National Natural Science Foundation of China (Grant No. 11774194, 11804182, 11861131008, 51831005), Beijing Natural Science Foundation (Grant No. Z190009), and the Beijing Advanced Innovation Center for Future Chip (ICFC). Work at NIMTE was supported by the National Natural Science Foundation of China (Grant No. 11874059), Key Research Program of Frontier Sciences, CAS, Grant NO. ZDBS-LY-7021, Beijing National Laboratory for Condensed Matter Physics, and Natural Science Foundation of Zhejiang Province (Grant No. LR19A040002). Financial support by the German Science Foundation is gratefully acknowledged within program SCHU 2992/4-1 and 2992/1-3. We thank Joseph Sklenar for his critical reading of our manuscript.

**Figures and Figure captions**

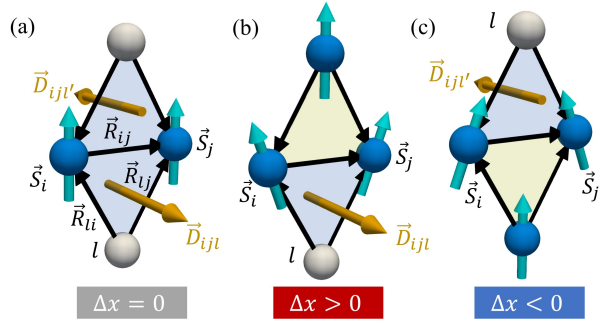

**Figure 1. The schematic diagram of the three-site Fert-Lévy model.** (a) The schematic illustration of the atom/spin structures without BMA, the DM vectors $\vec{D}_{ijl}$ and $\vec{D}_{ijl'}$ are described by the three-site Fert-Lévy model. The blue (gray) balls represent the FM and HM atoms. $\vec{R}_{li}$ and $\vec{R}_{lj}$ are the intersite vectors. (b) and (c) The asymmetric spin structure for systems with opposite BMA ($\pm\Delta x$) exhibit nonzero opposite DMI vectors.

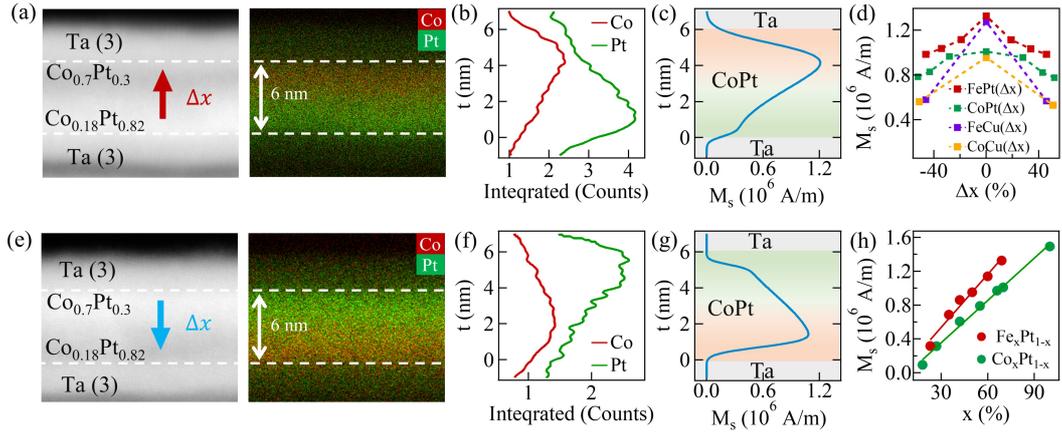

**Figure 2. Structural and magnetic evidence of BMA in CoPt ($\Delta x = \pm 52\%$).** The HAADF-STEM images of films with $\Delta x = +52\%$ (a) and $\Delta x = -52\%$ (e), respectively. For the $\Delta x = +52\%$ ($\Delta x = -52\%$), compositions from bottom to top change from $Co_{0.18}Pt_{0.82}$ ($Co_{0.7}Pt_{0.3}$) to $Co_{0.7}Pt_{0.3}$ ($Co_{0.18}Pt_{0.82}$), respectively. The opposite trends of contrast reveal an opposite element distribution of Co/Pt (red/green). (b) and (f) The EDS line profiles. (c) and (g) The magnetization depth profiles from the PNR measurement. (d) The dependence of $M_s$ vs. $\Delta x$ in CoPt ($\Delta x$), FePt ($\Delta x$), CoCu ($\Delta x$), and FeCu ($\Delta x$) films. (h) The evolution of $M_s$ vs. $x$ in films without gradient ($\Delta x = 0$).

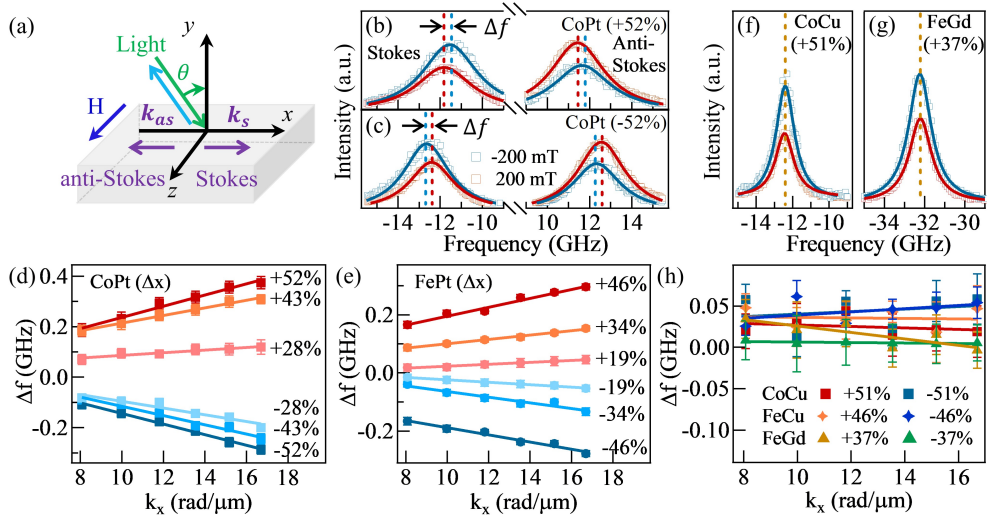

**Figure 3. The frequency shift of the nonreciprocal spin-wave propagation.** (a) The setup geometry of the BLS spectroscopy. The anti-Stokes/Stokes processes correspond to annihilation/creation of magnons. (b) and (c) The BLS spectra for CoPt ($\Delta x = \pm 52\%$) films in which the red (blue) curve represents the spectra with $\pm \mu_0 H$, respectively. (d) and (e) The summarized $\Delta f$ $vs.$ $k_x$ for the CoPt ($\pm \Delta x$) and FePt ($\pm \Delta x$) films. (f) and (g) The BLS spectra for CoCu ($\Delta x = +51\%$) and FeGd ($\Delta x = +37\%$). (h) The evolution of $\Delta f$ $vs.$ $k_x$ for the CoCu ($\pm \Delta x$), FeCu ($\pm \Delta x$) and FeGd ($\pm \Delta x$) films.

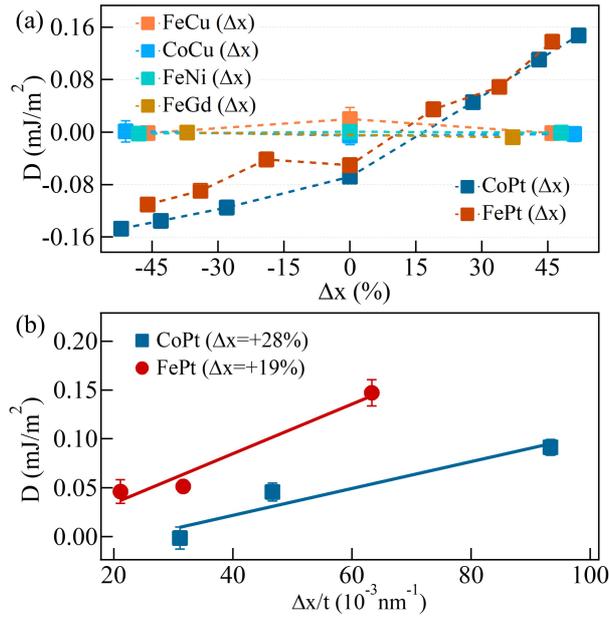

**Figure 4. The experimental results for the DMI induced by the BMA.** (a) The volume-averaged DMI constant $D$ for films with varied amplitudes/signs of $\Delta x$ ($t = 6$ nm). (b) The evolution of $D$ vs. $\Delta x/t$ for the CoPt ($\Delta x = +28\%$) and FePt ($\Delta x = +19\%$) films with different thicknesses. The solid line is a linear fitting curve.

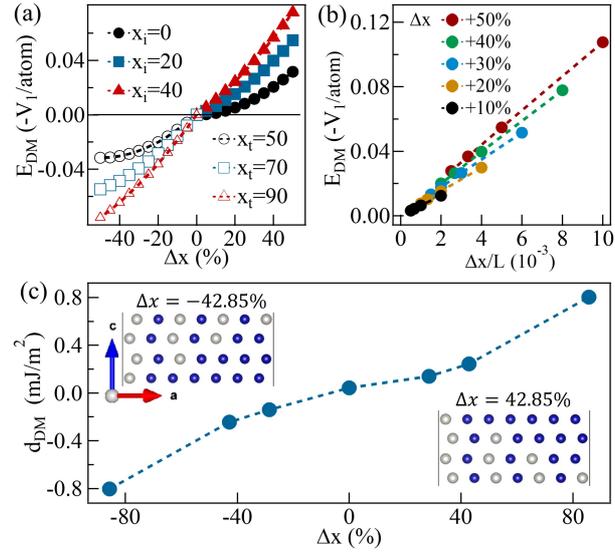

**Figure 5. The calculated DMI induced by the BMA.** (a) The calculated DMI energy $E_{DM}$ vs. $\Delta x$ with fixed starting/ending compositions ($x_{i,t}$). (b) The evolution of $E_{DM}$ vs. $\Delta x/L$. (c) The first-principles result of the gradient-dependent DMI constants ($d_{DM}$).

Cowburn, Three-dimensional nanomagnetism. *Nat. Commun.* **8**, 1 (2017).